\documentclass[12pt,onecolumn]{article}
\usepackage{graphicx}
\usepackage{fullpage}
\begin{document}

\begin{center}
{\bf \Large The growth and evolution of super massive black holes}
{\large A white paper submitted to the Decadal Survey Committee}
\end{center}
\vspace{0.2cm}
{\bf Authors:} S. Murray$^1$, R. Gilli$^2$, P. Tozzi$^3$, M. Paolillo$^4$, N. Brandt$^8$, G. Tagliaferri$^9$, A. Vikhlinin$^1$, M. Bautz$^{12}$, S. Allen$^{13}$, M. Donahue$^{14}$, K. Flanagan$^{15}$,
P. Rosati$^6$, S. Borgani$^5$, R. Giacconi$^{10}$, M. Weisskopf$^7$, A. Ptak$^{10}$, S. Gezari$^{10}$,
D. Alexander$^{11}$, G. Pareschi$^9$, W. Forman$^1$, C. Jones$^1$, R. Hickox$^1$\\
\noindent
{\footnotesize
1. Harvard-Smithsonian Center for Astrophysics, Cambridge MA\\
2. INAF-Osservatorio Astronomico di Bologna, Bologna, Italy\\
3. INAF-Osservatorio Astronomico di Trieste, Trieste, Italy\\
4. Universit\`a di Napoli, Napoli, Italy\\
5. University of Trieste, Trieste, Italy\\
6. European Southern Observatory (ESO), Garching bei Muenchen, Germany\\
7. NASA Marshall Space Flight Center, Huntsville AL\\
8. Penn State University, University Park PA\\
9. INAF-Osservatotio Astronomico di Brera, Milano, Italy\\
10. The Johns Hopkins University, Baltimore MD\\ 
11. University of Durham, Durham, United Kingdom\\
12. Massachuesetts Institute of Technology, Cambridge MA\\
13. Standford University, Stanford CA\\
14. Michigan State University, E. Lansing MI\\
15. Space Telescope Science Institute, Baltimore MD\\
}
\begin{center}
Science Frontier Panels\\
Primary Panel: Galaxies across Cosmic Time (GCT)\\
Secondary panel: Cosmology and Fundamental Physics (GFP)\\
\vspace{0.2cm}
Project emphasized: The Wide-Field X-Ray Telescope (WFXT);
http://wfxt.pha.jhu.edu/
\end{center}
\vspace{0.25in}

\begin{figure}[h]
\centerline{\includegraphics[height=0.42\linewidth]{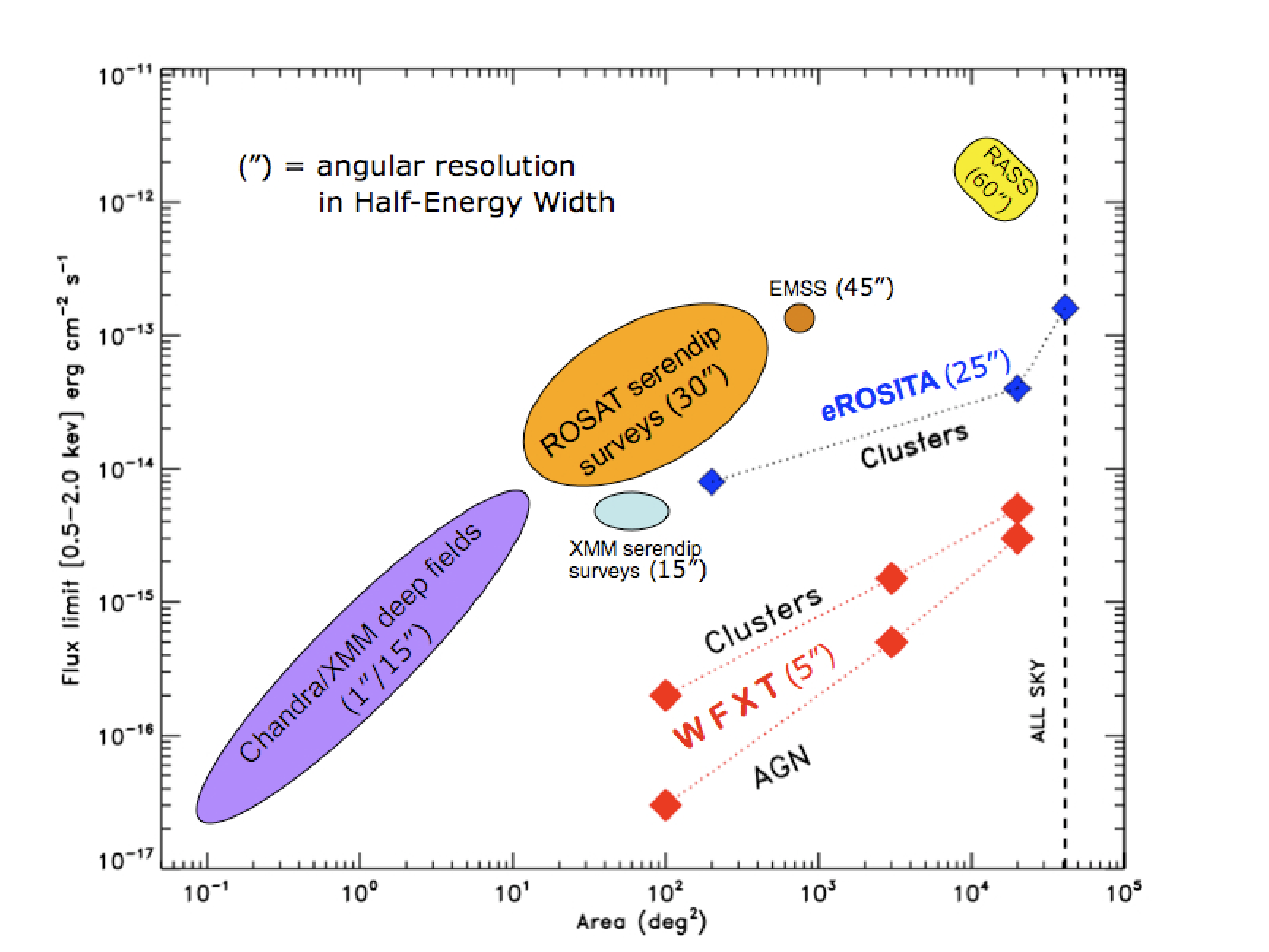}}
\end{figure}

\newpage
{\bf Executive summary.} We discuss the central role played by X-ray
studies to reconstruct the past history of formation and evolution of
supermassive Black Holes (BHs), and the role they played in shaping
the properties of their host galaxies. We shortly review the progress
in this field contributed by the current X-ray and multiwavelength
surveys. Then, we focus on the outstanding scientific questions that
have been opened by observations carried out in the last years and
that represent the legacy of Chandra and XMM, as for X-ray
observations, and the legacy of the SDSS, as for wide area surveys:
{\bf 1) When and how did the first supermassive black holes form? 2)
How does cosmic environment regulate nuclear activity (and star
formation) across cosmic time? 3) What is the history of nuclear
activity in a galaxy lifetime?} We show that the most efficient
observational strategy to address these questions is to carry out a
large-area X-ray survey, reaching a sensitivity comparable to that of
deep Chandra and XMM pointings, but extending over several thousands
of square degrees. Such a survey can only be carried out with a
Wide-Field X-ray Telescope (WFXT) with a high survey speed, due to
the combination of large field of view and large effective area, i.e.,
grasp, and sharp PSF. We emphasize the important synergies that WFXT
will have with a number of future ground-based and space telescopes,
covering from the radio to the X-ray bands and discuss the immense
legacy value that such a mission will have for extragalactic astronomy
at large.
\vspace*{-0.6cm}
\section{Introduction}
\vspace*{-0.4cm}
Several pieces of evidence point towards an intimate relation between
the evolution of galaxies and the accretion of supermassive black
holes (SMBHs) at their centers, indicating that most galaxies in the
Universe spent a fraction of their lifetimes as active galactic nuclei
(AGN). In the local Universe, most galaxy bulges host a SMBH
\cite{ff}, whose mass scales with the bulge mass and stellar velocity
dispersion \cite{fm,gebhardt}. Furthermore, the growth of SMBHs during
active accretion phases, traced by the cosmological evolution of the
AGN luminosity function \cite{ueda,hasinger,lafranca,silverman},
eventually matches the mass function of SMBHs in the local Universe
\cite{marconi,yu,shankar}. Finally, both BH growth and star formation
appear to follow the same "anti-hierarchical" behaviour over cosmic
time. The peak activity of luminous QSOs occurs at $z~\sim 2$, where
large galaxies were also forming most of their stars, while moderately
luminous AGN are more common at the current epoch, where stars are
forming is smaller galaxies. The role played by nuclear activity in
modulating star formation processes in the host galaxies
(``feedback'') is the subject of huge research efforts.

While the SMBH vs galaxy co-evolution is now an accepted scenario, the
details of this joint evolution are not yet fully understood. In
particular, the processes that control BH growth are uncertain.
Nuclear activity in bright QSOs is thought to be induced by major
mergers or close encounters of gas-rich galaxies in the context of
hierarchical structure formation \cite{cv,hopkins}, but the case
for the importance of such mergers in low-to-moderate luminosity AGNs
is much less clear \cite{grogi}.
Furthermore, the early phases of nuclear activity triggered by merger
events are thought to be buried by large columns of gas and dust,
possibly with concurrent vigorous star formation
\cite{hopkins}. Evidence for coeval star formation and nuclear
activity produced by galaxy interactions has been indeed found in
populations of bright IR/submillimeter galaxies at $z\sim2$ and beyond
\cite{alex}. In contrast, the majority of $z\sim 2$ AGN selected at
faint X-ray fluxes are hosted by galaxies with a spectral energy
distribution typical of passively evolving objects \cite{mainieri}. It
has been suggested that the concurrent growth of black holes and
stellar mass observed in IR galaxies at $z\sim 2$ is a long-lived
($>0.2$ Gyr) phenomenon, unlikely to be triggered by rapid
major-merger events \cite{daddi}, which is instead the common
interpretation to explain the highly efficient star formation and BH
accretion observed in the highest redshift $(z>6)$ QSOs to date
\cite{li}.  The current picture has several unsolved issues:\\ 1) the
early stages of BH and galaxy formation, where strong star formation
events are expected to be associated with vigorous nuclear accretion,
are still unknown.  \\ 2) the dependence of the cosmic accretion
history on galaxy merging and interactions, hence on galaxy
environment, is currently debated.  \\ 3) the history and duty cycle
of nuclear activity in a galaxy lifetime, especially the relative
phases of obscured and unobscured accretion, have still to be
understood.

Since most of the accretion onto SMBHs is expected to be obscured,
deep X-ray surveys complemented by multiwavelength followup are needed
to sample the whole AGN population. At the moment such surveys
(e.g. CDFs, COSMOS, AEGIS) do not have enough sensitivity nor sky
coverage to provide the statistics necessary to address the above
issues.
\vspace*{-0.6cm}
\section{Super Massive Black Holes in the Early Universe}
\vspace*{-0.4cm}
{\bf When and how did the first supermassive black holes form?}\\
While most stars in the Universe appear to have formed at $0.5<z<3$,
when SMBHs were also growing most of their mass \cite{heck}, the first
objects formed at even earlier epochs, as soon as baryons were able to
cool within dark matter halos. To date, about 30 galaxies and about 20
QSOs have been observed at redshifts above 6\cite{tani,fan}. The BH
masses measured for these very high redshift QSOs are of the order of
$10^9\;M_{\odot}$ \cite{kurk}, which must have been built in less than
1 Gyr (the age of the Universe at $z=6$). These giant black holes are
thought to have formed through mass accretion onto smaller seed black
holes with mass in the range $10^2-10^4\; M_\odot$. A number of
possibilities for the origin of these seed black holes have been
proposed, which range from being the $10^2\; M_\odot$ remnants of
massive, PopIII, stars \cite{madau}, to being the $10^4\; M_\odot$
products of direct collapse of large molecular clouds
\cite{volo}. Whatever the seed origin is, the huge mass gain over a
cosmologically short timescale implies that the BH growth in these
objects was close to (or even higher than) the Eddington limit and
continuous over that time. Recent hydrodynamical simulations, have
shown that, within large dark matter halos, merging between
proto-galaxies at $z\sim 14$ with seed BH masses of $10^4\; M_\odot$
may trigger Eddington limited nuclear activity to produce a $10^9$
solar mass BH by $z\sim 6$ \cite{li}. The frequency and efficiency
with which this merging and fueling mechanism can work is however
unknown.

Currently, given the scarcity of observational constraints, BH and
galaxy formation at early epochs is speculative work. For example, the
relation between the BH mass and the galaxy gas mass has been measured
for a few bright QSOs at $z\sim 5-6$, in which the ratio between the
BH mass and that of the dynamical mass within a few kpc radius, as
measured from CO line observations, is of the order of 0.02-0.1
\cite{walter,maio}. This is more than one order of magnitude larger
than the ratio between the BH mass and stellar bulge mass measured in
local galaxies (see also \cite{peng} for similar conclusions for at a
sample of $z>1$ AGN). Despite the uncertainties on these measurements
(see e.g. \cite{narayanan}), the possibility that SMBHs are leading
the formation of the bulge in proto-galaxies, poses severe constraints
to galaxy formation models. In general, it is not clear whether the
$M_{BH}/M_{bulge}$ ratio measured in these objects is a feature of
peculiar sources or it can be extrapolated to the full galaxy
population (see e.g. \cite{alex08}). Large statistical
samples of $z>6$ AGN are needed to understand the relation between the
BH growth and formation of stars in galaxies at their birth.

Another observational constraint regarding accreting SMBHs in the
early Universe is the space density measured for luminous,
optically-bright QSOs selected by the SDSS, which suggests that the
the space density of quasars with $L_{bol}>10^{46}$ erg/s declines
from $z\sim 3$ to $z>6$ \cite{fan}. This measurement only applies to
the most luminous optical QSOs, i.e. to a presumably tiny fraction of
the active BH population at $z>6$, which is instead expected to be
made primarily by less massive, $10^6\;M_{\odot}$, and less luminous,
$L_{bol}\sim10^{44}$ erg/s, objects. Many semi-analytical models of
early BH formation within the growth of cosmic structures have been
proposed in recent years \cite{marulli,salvaterra,rhook}, in which the
BH formation rate depends on several factors such as: i) where
(i.e. in which dark matter halos) they form; ii) what is their average
accretion rate; iii) what is the triggering mechanism: galaxy merging
and fly-by?  Accretion of cold gas independent of galaxy interactions?
Measuring the space density of low luminosity AGN at $z>6$ would
constrain most the model parameters, and will therefore be the key to
understanding how the population of early BHs has formed.

While optically-bright QSOs at high redshift can be efficiently
selected by optical color techniques (e.g. by wide-area optical
surveys with LSST and Pan-STARRS) and then identified
spectroscopically, this is not the case for obscured high-z QSOs,
since they would escape standard color selection. Furthermore,
high-excitation narrow emission lines -- the sign of nuclear activity
in the optical/near-IR spectra of obscured AGN -- are often weak or
absent. Therefore, major future facilities like JWST or ALMA will
measure the redshift of galaxies up to $z>8$, but will not reliably
establish if an obscured AGN hides at their centers.

\begin{figure}[t]
\includegraphics[width=6cm]{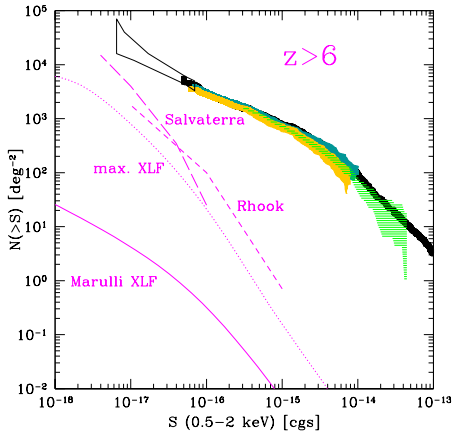}
\begin{minipage}{9cm}
\vspace{-6cm}
\caption{\small The 0.5-2 keV logN-logS expected for AGN at $z>6$
according to different models as labeled. Note the vertical scale and
the very large scatter in the model predictions. The detection with
WFXT of a few thousand objects at $z>6$ (and $\sim$ a hundered at
$z>8$) at fluxes of $10^{-17}$-$10^{-16}$ erg/cm$^2$/s would constrain
early BH formation. We note that such a practical goal is beyond the
capabilities of eROSITA, which is confusion limited around fluxes of a
few $\times 10^{-15}$ erg/cm$^2$/s, where the surface density of
high-z AGN is extremely low.}
\label{sky}
\end{minipage}
\vspace{-0.5cm}
\end{figure}

To date, {\bf no} obscured AGN at $z>6$ have been discovered, as
opposed to 20 unobscured QSO at $z>6$. Nonetheless, there are several
arguments which suggest that a large population of obscured QSOs,
probably even larger than that of unobscured QSOs, has to be present
at high redshift. First, observations and modeling up to $z<4$ have
shown that obscured AGN outnumber unobscured ones by large factors
(2-8) \cite{gilli} which have been also proposed to increase with
redshift \cite{treister}. Second, current models of galaxy formation,
postulate that the early phase of accretion onto the central black
hole is obscured, which would naturally predict many high-z obscured
sources. Third, dust and metals have been found to be abundant in the
inner regions of $z>6$ objects \cite{yuarez,beelen}, which could
therefore absorb nuclear radiation in the optical/UV band. The lack of
known obscured AGN at very high-redshift is presumably related to the
limitations of current sky surveys. Indeed, the large volumes needed
to detect rare, high-z objects have been achieved only by optical
surveys like the SDSS and UKIDSS, which are bound to optical color
selection criteria, while deep X-ray surveys, which efficiently select
obscured objects, are currently covering too small a volume. The
discovery space for early obscured AGN is therefore huge, and can have
an extreme impact on our understanding of BH and galaxy formation. All
Sky Surveys like the RASS are too shallow to detect faint, obscured
AGN at $z>6$, and the X-ray surveys to be performed by the near future
mission eROSITA will be confusion limited at X-ray fluxes too bright
to detect $z>6$ objects (see Fig.~1 and 3).

This discovery space can be filled through deep X-ray surveys over
large sky areas. To this end an X-ray facility with sharp ($\sim 5''$
HPD) resolution constant over a large ($\sim 1$ deg$^2$) FOV, and
large ($\sim 0.5-1.0$ m$^2$ at 1 keV) effective area would be
ideal.\footnote{It is recalled that, for objects at $z>6$, photons
observed at 1 keV correspond to rest frame energies of $>7$ keV,
i.e. they will be largely unaffected by nuclear obscuration.} The
above mentioned requirements are planned for the WFXT mission
\cite{murray08}. In 5 years WFXT can easily observe the equivalent of
a thousand Chandra Deep Fields\footnote{We note here that, when
averaged over the field of view, the resolution of Chandra and WFXT
will be comparable, Chandra being significantly better that WFXT on
axis but significantly worse off axis.}, plus a few thousand
Chandra-COSMOS fields, plus several thousand XBo\"otes fields. Based
on extrapolations of current X-ray luminosity functions at $z<4$, in
which the AGN space density declines exponentially at high redshifts
(see e.g. \cite{brusa}) one would expected to detect 1000 obscured
AGN, plus 1300 unobscured AGN at $z>6$. Also, approximately a hundred
AGN are expected at $z>8$. While optical surveys with LSST and
Pan-STARRS may be used to isolate candidates high-z unobscured AGN
among WFXT sources, the identification process for obscured objects is
more problematic and will need a joint effort with the major future
optical to IR facilities. Deep and wide photometric IR surveys such as
those planned with VISTA and spectroscopic IR surveys like those under
study for the ESA/NASA JDEM/Euclid project, are expected to provide
redshifts for faint, high-z galaxies over a large portion of the
sky. A match between an apparently quiescent galaxy at $z>6$ revealed
by optical/near IR surveys and an X-ray source detected by WFXT would
reveal the presence of a hidden AGN in the early Universe.

It is worth noting that the above numbers of high-z AGN to be observed
with WFXT are reference numbers. Indeed, the predictions based on
semi-analytic models scatter by several orders of magnitude depending
on the model assumptions, and it is immediately clear that a
statistically large sample such as that provided by WFXT is needed to
constrain models of QSO and galaxy formation. Even simple comparisons
between the observed source counts at $z>6$ (and at $z>8$) with model
predictions will do the job (see Fig.~1). WFXT will then constitute a
unique tool to probe the population of high-z obscured AGN, a
scientific issue which is not achievable with any of the planned
facilities in the next 20 years. Even the proposed International X-ray
Observatory (IXO), which will have the power to get good quality X-ray
spectra for faint high-z sources, will be limited by its small FOV. 
In this regard, there will be full complementarity, with WFXT detecting
high-z obscured candidates to be followed up by IXO.
\vspace*{-0.6cm}
\section{Black hole fueling and the Large Scale Structure}
\vspace*{-0.4cm}
{\bf How does cosmic environment regulate nuclear activity and star
formation?}\\ The role played by the environment
- voids, filaments, groups, clusters - in triggering both nuclear
activity and star formation is still a matter of debate. 
\begin{figure}
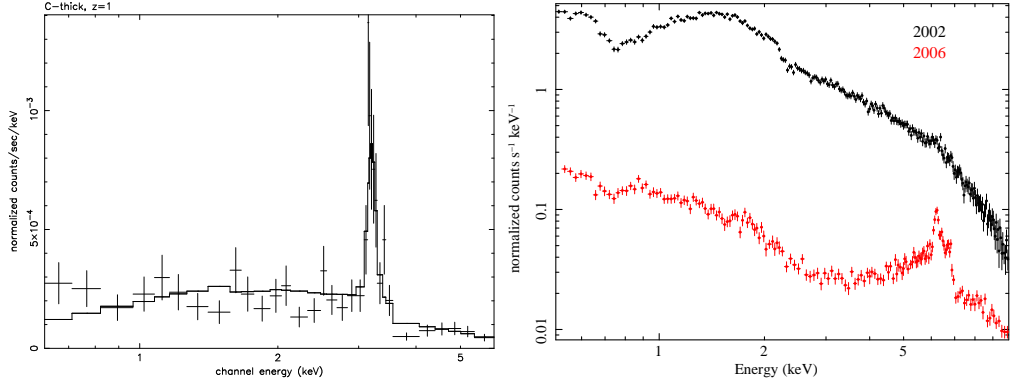

\begin{center}
\includegraphics[width=5cm,angle=270]{z1_thick_fh4m15_400ks_g10_lin.ps}
\includegraphics[width=5cm,angle=270]{lon.ps}
\caption{\small {\it Left Panel}: simulated X-ray spectrum of a
moderately bright Compton-thick AGN at $z=1$ observed with WFXT. About
500 such objects are expected to be observed with WFXT in 5
years. Note the prominent iron K$\alpha$ line which is both the
hallmark of a Compton-thick nucleus and provides the source redshift
independently of optical follow-up. {\it Right panel}: the XMM
spectrum of the AGN H0557-385 observed at two different
epochs, which is indicative of extremely variable X-ray
absorption \cite{longi}. Only $\sim 10$ such objects have been observed to
date. The discovery space that will be opened by WFXT on this subject
is huge.}
\label{spec}
\end{center}
\vspace{-0.7cm}
\end{figure}
The major deep-and-wide multiwavelength surveys to date, like COSMOS
and AEGIS, are producing the first attempts to build matter density
maps over a relatively large redshift range, and study the fraction of
active galaxies as a function of matter density
\cite{silverman}. Given the limited volumes ($10^6-10^7$ Mpc$^3$)
covered, these surveys cannot span the full environment range
(e.g. rare, massive galaxy clusters are not sampled) and are dealing
with limited AGN statistics. Current results are therefore still weak.
Increasing the object statistics and having the ability to identify
low level AGN activity associated with galaxies in several
environments is needed. Clustering techniques can be also used to
study the relation between nuclear activity and the environment. The
comparison between the clustering properties of AGN and those of dark
matter halos predicted by cold dark matter models are commonly used to
evaluate the typical mass of the dark matter halos in which AGN form
and reside as a function of cosmic time (e.g. $M>10^{12} \: M_{\odot}$
for bright optical QSOs at $z<4$ \cite{porciani,croom05}). The ratio
between the AGN space density and that of host dark matter halos
provides an estimate of the AGN lifetime \cite{mw}, which, based on
the results from optical surveys, is loosely constrained in the range
$10^6 - 10^8$ yr \cite{grazi04,porciani}. Very recent works based on
X-ray selection suggest instead lifetimes as long as 1 Gyr for
moderately luminous AGN. The comparison between the clustering
properties of different galaxy types and AGN can be used to estimate
AGN hosts and to estimate the descendant and progenitors of AGN at a
given redshifts.

The SDSS and the 2QZ surveys have measured the clustering of bright
QSOs up to $z=4$. However, they are not sampling the bulk of the
nuclear accretion in the Universe, since they are limited to very
luminous QSOs $(L_{bol}> 10^{46} L_\odot)$. Only sparse clustering
measurements have been performed using the major multiwavelength
surveys to date \cite{gilli09, coil, hick}.

A very large - a few millions - AGN sample selected in the X-rays,
which can be divided into several luminosity, redshift, obscuration,
and environment bins, will be an ideal tool to study the relation
between nuclear activity and galaxy interactions, since it will allow
to measure the AGN luminosity function and evolution as a function of
cosmic environment. Such a large sample will measure AGN clustering as
a function of redshift, luminosity and obscuration.

The WFXT mission is expected to provide a $10^7$ AGN sample
distributed over the entire AGN luminosity vs redshift plane. This
sample will also include $\sim10^6$ heavily obscured AGN ($N_H>10^{23}
cm^{-2}$), allowing studies of the evolution of AGN obscuration as a
function of redshift and environment. In addition, it will return a
sizable sample of the most obscured ($N_H>10^{24} cm^{-2}$),
Compton-thick AGN at $z\geq 1$, i.e. the missing source population
expected to produce $\sim 1/4$ of the energy density of the cosmic
X-ray background \cite{gilli,treister}. The typical optical
counterparts of moderately bright X-ray AGN are expected to have
$R\sim22-23$ mag and will be accessible by most current and planned
large area surveys (VISTA, LSST, Pan-STARRS). Notably, for about $500$
Compton-thick AGN at $z\geq1$ (i.e. the brightest tail), a good
quality X-ray spectrum will be obtained (see Fig.~2 left). For these
objects, the prominent iron line at 6.4 keV, shifted to $\sim 3$ keV
in the observed frame, can be used as a powerful tool for both
identifying the source as Compton-thick and measuring its redshift
independently of any optical identification process. Such a large
sample of {\it bona-fide} Compton-thick AGN at $z\geq1$ is beyond
reach of currently proposed hard X-ray missions (see e.g. Fig.~3).  As
an example, projects like EXIST and NuSTAR, which are less sensitive
than WFXT, will basically detect moderate redshifts objects ($z<0.5$
or so), i.e. they will not sample the Compton-thick population
responsible for the missing CXB. NeXT and Simbol-X instead, will be
able to detect $z\geq1$ Compton-thick AGN, but because of the small
grasp grasp, their object statistics will be very limited.

\begin{figure}
\includegraphics[height=6cm,angle=0]{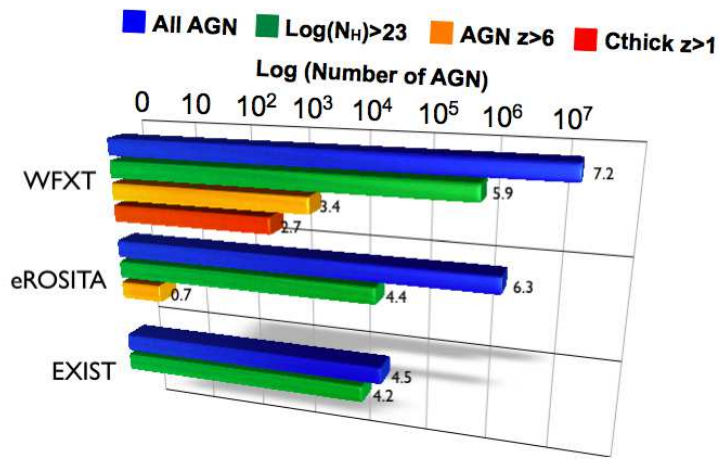}
\label{histo}
\begin{minipage}{6cm}
\vspace{-6cm}
\caption{\small Comparison between the capabilities of future/proposed X-ray
missions in providing statistically large samples of objects:
blue=total number of AGN detected; green=number of obscured AGN with
$N_H>10^{23}$; yellow=number of AGN detected at $z>6$; orange=number
of {\it bona fide} Compton thick AGN at $z\geq1$, i.e. those for which
a spectrum of the quality shown in Fig.~2 $left$ will be obtained.  }
\end{minipage}
\vspace{-0.5cm}
\end{figure}
\vspace*{-0.5cm}
\section{AGN variability}
\vspace*{-0.5cm}
{\bf What is the history of nuclear activity in a galaxy lifetime?}\\
AGN activity manifests itself through intense variability across the
entire electromagnetic spectrum on time-scales from minutes to
years,
which is an excellent probe of the physical conditions within the most
energetic, inner regions. It is now believed that the time dependence
of the SMBH emission is closely linked to the underlying physical
processes and that these are similar to those characterizing BH
accretion in galactic binaries \cite{u97}.

The variability properties of SMBH have been studied in detail only
for a handful of nearby sources, for which long monitoring campaigns
and high fluxes make it possible to put constraints on the physics of
accretion. The best measurements to date suggest
that variability time-scales scale with BH mass and accretion rate
\cite{Ut02, Mk03, McH04}. A few sources have been observed
to change between a weakly or a heavily obscured state (see Fig.~2
right), for which the structure and location of the absorber have been
studied in detail \cite{Ris07,Ris09}. For a few other objects, flux
and spectral variability over the years trace the on/off switching of
the nuclear activity \cite{P04}.

Little is known about the variability properties of AGNs at
intermediate to high-redshifts, due to the lack of large grasp
monitoring surveys. However, the few studies based on the monitoring
byproduct of deep X-ray surveys (CDFs and Lockman-Hole) verified that,
when observed with sufficient photon statistics, most AGNs vary over
timescales from hours to years \cite{Mann02, P04, Pap08}. These studies
also suggested that variability may have been more extreme in the past
($z\geq 1$), possibly due to an increase in accretion rates with
lookback time.
If the link between X-ray
variability with mass and accretion rate, recently discovered for
local AGNs, is confirmed, monitoring distant AGNs through high-energy
observations will provide an unique tool to infer the mass and
accretion properties of the SMBH population.

Transient X-ray outbursts from galactic nuclei are also expected when
a star, planet, or gas cloud is tidally disrupted and partially
accreted by the black hole. A few candidates were discovered in the
X-rays by comparing multi-epoch ROSAT observations which detected
large amplitude flares (factors of $\sim 20-400$ or more) from
non-active galaxies, with large peak X-ray luminosities ($10^{42-44}$
erg s$^-1$), a decay timescale of months, and a measured detection
rate of 10$^{-5}$ yr$^{-1}$\cite{donley}. The detection of significant
samples of transient nuclei will be a new constraints to models of
stellar dynamics at the galaxy centers.

The above variability studies primarily call for an X-ray survey
mission with large grasp.
By performing repeated passes over the surveyed area, the WFXT mission
will provide the largest AGN sample specifically designed for
variability studies. The number of both temporally varying AGNs and
transient outbursts in galactic nuclei should increase by orders of
magnitude providing sufficient fundamental statistics to understand
the AGN inner structure. WFXT will be able to observe variability in
$10^5$ AGNs in a 100 deg$^2$ survey down to $\Delta f/f\sim 20\%$, and
reconstruct mass and accretion rates for several thousand sources.

Future wide area optical surveys (LSST, Pan-STARRS) will effectively
open the time domain to cosmological studies. The synergy with a wide
area X-ray mission with monitoring capabilities is expected to open up
a new horizon to variability studies.

\vspace{-0.5cm}
{\scriptsize
{}

\end{document}